\documentclass[letterpaper,journal]{IEEEtran}

\usepackage{cite}
\usepackage{amsmath,amssymb,amsfonts}
\usepackage{array}
\usepackage{booktabs}
\usepackage{balance}
\usepackage{graphicx}
\usepackage{stfloats}
\usepackage{textcomp}
\usepackage{xcolor}
\usepackage{url}

\AtBeginDocument{\setcounter{dbltopnumber}{1}}

\graphicspath{{figures/}}
\begin{document}

\title{Reassessing 3GPP NR CSI Codebook Structures in Near-Field Channels: Finite-Feedback Multilayer Precoding and Design Insights}

\author{Hongbo Xing, Jianhua Zhang, Yuxiang Zhang, Huixin Xu, Haiyang Miao, and Lei Tian%
\thanks{H. Xing, J. Zhang, Y. Zhang, and L. Tian are with the State Key Laboratory of Networking and Switching Technology, Beijing University of Posts and Telecommunications, Beijing 100876, China (e-mail: hbxing@bupt.edu.cn; jhzhang@bupt.edu.cn; zhangyx@bupt.edu.cn; tianlbupt@bupt.edu.cn). H. Xu is with Peking University Shenzhen Graduate School, Shenzhen 518055, China (e-mail: xuhuixin@pku.edu.cn). H. Miao is with the State Key Laboratory of Intelligent Technologies and Systems, Department of Automation, Tsinghua University, Beijing 100084, China (e-mail: miaohy@tsinghua.edu.cn). (Corresponding author: Yuxiang Zhang.)}}

\maketitle

\begin{abstract}
3GPP TR 38.901 Rel-19 introduces antenna-element-level spherical-wave modeling, while NR Type-I and enhanced Type-II (eType-II) CSI codebooks continue to use plane-wave DFT beams. Whether this mismatch materially degrades finite-feedback multilayer precoding in standardized multipath channels remains unclear. To isolate its impact, we evaluate both codebooks over strictly paired far-field (FF) and near-field (NF) 3GPP channels that share user locations, multipath parameters, polarization, and link budgets and differ only in their wavefront models. Simulations cover Rank 1--4 transmission in 7-GHz UMi and 24-GHz InH-linear scenarios. We find no systematic FF/NF shift in singular-mode gains or equal-power SVD (SVD-EP) rates. Instead, spherical-wave phases reorder multipath projections onto plane-wave candidates and thereby alter codeword selection. For Rank-4 InH-linear users at 0.1 normalized Rayleigh distance, given FF-selected Type-I and eType-II codewords incur median direct mismatch losses of 2.28\% and 5.34\% on the NF channel, respectively; codebook reselection identifies better-matched codewords and reduces these losses to 0.770\% and 3.28\%. Nested candidate-set comparisons further show that relaxing Type-I interlayer constraints improves the SVD-EP-normalized rate by 20.7 percentage points, whereas finite-range sampling adds only 0.579 points. These results support prioritizing multilayer multibeam representation in large-aperture NR CSI codebooks, with range states providing complementary refinement.
\end{abstract}


\section{Introduction}
\label{sec:introduction}

\subsection{Background and Motivation}

Large-aperture extremely large-scale multiple-input multiple-output (XL-MIMO) arrays improve angular resolution~\cite{unified6gchannel} and can exploit the additional spatial degrees of freedom in three-dimensional channels to enhance spatial multiplexing gains~\cite{zhang2017threedmimo}; they have therefore become an important candidate technology for sixth-generation mobile systems~\cite{tataria2021sixg}. At the implementation level, integrated millimeter-wave full-digital beamforming arrays have been developed for B5G/6G operation~\cite{lin2024fullDigitalArray}, while OTA-calibrated phased arrays have shown practical per-channel phase control and beam steering near 26~GHz~\cite{wu2024otaPhasedArray}. As apertures grow and operating wavelengths shorten, more cellular users and scatterers enter the radiative near field; distance-dependent array responses caused by spherical-wave propagation have already been observed in channel measurements~\cite{sub6mmwave2023}. To represent these propagation characteristics in cellular scenarios, 3GPP TR~38.901 Release~19 introduced element-level near-field propagation and corresponding calibration configurations into the 7--24-GHz channel model~\cite{3gpp38901Rel19,zhang2025rcs}; its propagation mechanism, parameterization, and calibration procedure have been documented systematically~\cite{xuhuixin-3GPP-channel}. Meanwhile, the NR Type-I and eType-II CSI codebooks retain plane-wave DFT bases as their fundamental spatial structure~\cite{3gpp38214Rel19}. Their ability to represent a multi-layer channel therefore merits reassessment when spherical-wave propagation is introduced into standardized multipath channels.

\subsection{Related Work}

Spherical-wave propagation changes the channel in two principal ways. First, it changes the phase structure of an individual path across the array aperture. A far-field response has a phase that is linear in element position and can be represented by a DFT beam at one direction. A near-field response is determined by the exact element-to-user or element-to-scatterer distance and consequently contains phase curvature governed by direction, range, and aperture~\cite{cui2022polar}. When a fixed far-field beam is applied to a near-field path, the element phases no longer combine fully coherently and the beamforming gain decreases; this angle--range coupling is a defining distinction between near-field focusing and far-field directional scanning~\cite{bjornson2021primer}. A spherical-wave response also spreads from a single spatial frequency into a continuous plane-wave angular spectrum~\cite{pizzo2022fourier}. A finite array represents this spectrum with multiple DFT frequency bins whose occupied range depends on direction, range, and aperture~\cite{kosasih2025spatial}. This DFT representation can be viewed as discrete sampling of the continuous wave-number spectrum, whose spread interval also admits a closed-form approximation~\cite{xing2025wavenumber}.

Polar-domain dictionaries jointly sample angle and range to construct near-field focusing vectors~\cite{cui2022polar} and have been applied to near-field channel estimation and beam training~\cite{abdallah2025polar}. Two-stage search reduces the overhead of angle--range beam training~\cite{zhang2022fastbeam}, while spatial-chirp beams enable hierarchical training codebooks~\cite{shi2024spatialchirp}; unified codebooks can also cover both far- and near-field users~\cite{zhang2024codebook}. In addition to explicit range states, a weighted combination of several DFT plane waves can approximate the spherical-wave response at a prescribed focal point~\cite{wang2024pwe}. Position-domain focusing in multi-user transmission can further separate users with similar directions but different ranges and suppress mutual interference~\cite{zhang2022beamfocusing}.

Second, spherical-wave phases also change the spatial modes of the complete MIMO channel. Curvature over a large aperture makes inter-element spatial correlation distance dependent~\cite{dong2022spatialcorrelation}; in a multipath channel, element-level phase changes further alter coherent path superposition and redistribute the gains of existing modes according to user geometry. When both link ends have large apertures, LoS aperture coupling may also create additional resolvable modes~\cite{ruizsicilia2025spatial}, affecting stream selection, RF-chain activation, and precoding~\cite{wu2022distanceaware} as well as double-sided near-field processing~\cite{shi2025doublesided}. Such modal growth is jointly limited by the effective projected apertures, wavelength, and propagation distance, and a small UE array substantially compresses the exploitable near-field capacity gain~\cite{miao2025capacity}.

\begin{figure*}[!t]
    \centering
    \includegraphics[width=\textwidth]{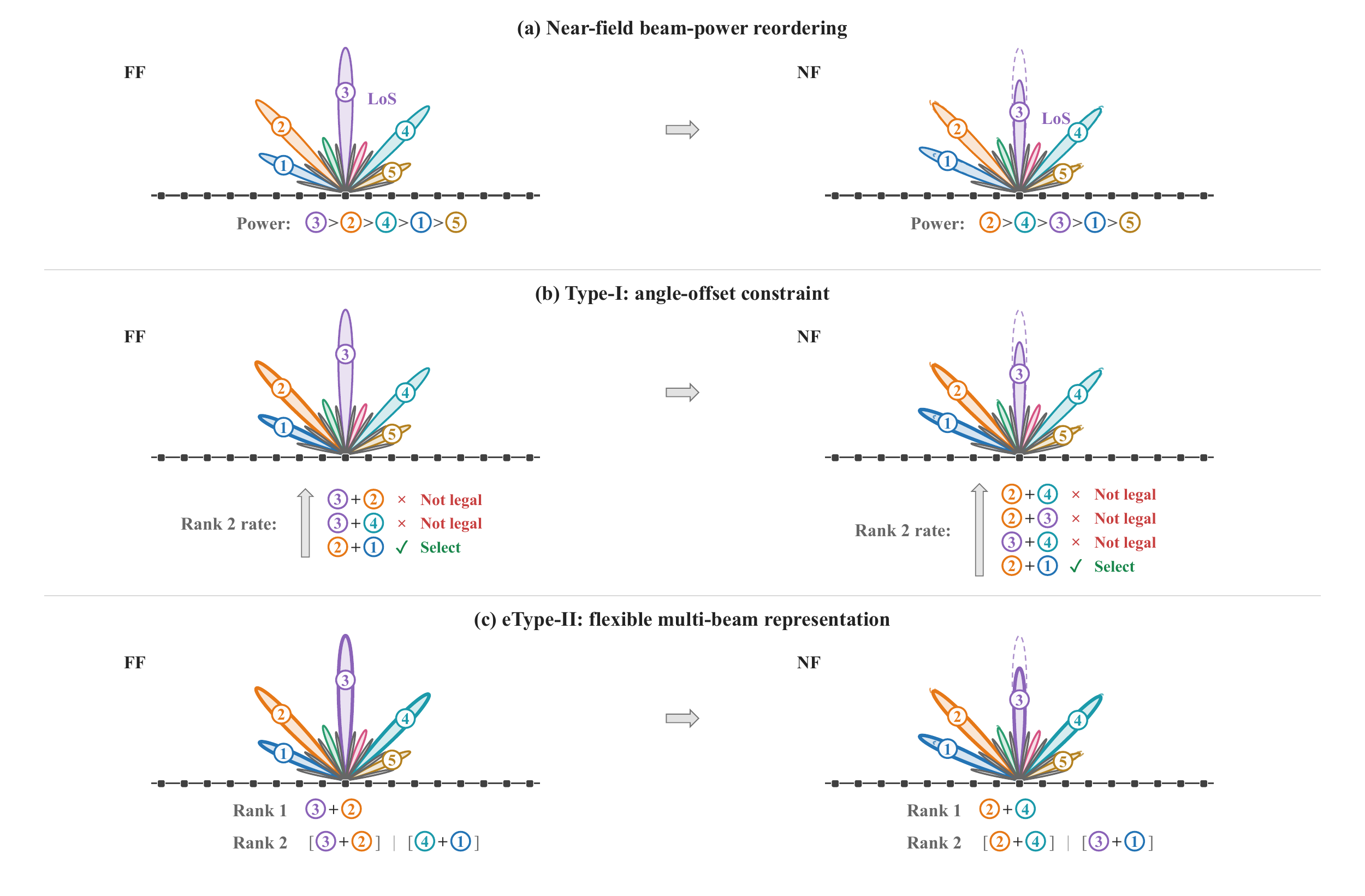}
    \caption{Candidate-beam gain reordering and finite-codebook selection under spherical-wave propagation: (a) nonuniform LoS/NLoS gain changes, (b) Type-I angle-offset constraints, and (c) flexible eType-II multibeam recombination.}
    \label{fig:beam-reordering}
\end{figure*}

\subsection{Problem Statement and Scope}

The central problem is how element-level spherical-wave propagation changes finite-feedback multilayer precoding with NR CSI codebooks. The two effects appear at the levels of a prescribed codeword and the complete channel, respectively. Spherical-wave phases change both the coherent gain of a fixed plane-wave codeword on an individual path and the superposition of multipath components in the complete MIMO channel. They thereby redistribute the relative gains of existing modes and strengthen or weaken different users' beam projections by different amounts. In the large-BS-aperture, small-UE-aperture cellular links considered here, clustered multipath and dual-polarized components already form multiple spatial modes, while the finite UE aperture further limits LoS modal growth. Precoding must also support rank-1--4 transmission under finite-feedback codebook constraints. In TS~38.214, Type-I constructs multilayer codewords from a small number of DFT beams, rank-dependent offsets, and polarization co-phasing, whereas eType-II forms a precoding matrix from several spatial basis beams and quantized complex coefficients~\cite{3gpp38214Rel19}. The two structures therefore provide different multilayer spatial-representation capabilities and feedback characteristics~\cite{fu2023precoder}. Existing codebook studies have also developed Type-II support-set and coefficient-feedback enhancements for UE mobility~\cite{ramireddy2022typeii}. Single-stream beam gain and rate under FF and NF models have been compared in a 3GPP scenario~\cite{ding2025farfield}, but single-path beam matching, ideal LoS MIMO, and single-stream evaluation do not determine the final effect of spherical-wave propagation on finite-feedback multilayer precoding over a dual-polarized multipath channel.

We therefore address two questions. First, over paired 3GPP far-field (FF) and near-field (NF) channels, how does spherical-wave propagation change the gain relations among existing MIMO modes, the projection gains and ordering of plane-wave candidate beams, and the multilayer rates of Type-I and eType-II? Second, when extending an existing codebook to near-field channels, should the design prioritize explicit range states or stronger multilayer spatial combinations, multibeam selection, and complex-weight representation?

Fig.~\ref{fig:beam-reordering} summarizes the chain linking spherical-wave phase, coherent multipath superposition, plane-wave projection reordering, and finite-codebook selection. After paired FF/NF channels are projected onto the same plane-wave candidates, the spherical-wave phase weakens the LoS-dominated direction while other strong multipath directions may strengthen or weaken, thereby changing the candidates' relative ordering. Type-I rank-dependent offsets may prevent several strong directions from entering the same codeword, so the best legal codeword can use candidates with lower individual projection gains; eType-II can instead recombine these directions through shared basis beams and per-layer complex weights. The final rate depends jointly on wavefront-induced projection reordering and the multilayer spatial combinations admitted by the codebook.

For each user, we generate strictly paired FF/NF channels that share location, cluster and ray parameters, polarization, path loss, and noise conditions and differ only in the element-level wavefront model. LoS aperture--range scaling and single-codeword coherent gain first characterize modal growth and beam mismatch. We then compare the singular modes, SVD-EP rates, and absolute Type-I/eType-II rates of the paired channels after extrapolating their port dimensions. Applying the FF-selected codeword directly to the corresponding NF channel and then comparing it with the rate after NF-based reselection separates the prescribed-codeword response from the portion recovered by codebook adaptation. Finally, Type-I, free-plane-wave, and polar combinations separate multilayer structural gain from range-sampling gain, while an eType-II multibeam extension provides an independent control without explicit range states.

\subsection{Contributions}

The main contributions are summarized as follows.
\begin{itemize}
    \item We develop a paired-channel evaluation framework for finite-feedback multilayer precoding with NR CSI codebooks. The FF/NF channels switch the element-level wavefront while preserving user location, path parameters, polarization, and link budget, enabling single-path geometry, channel-mode variation, and codebook-rate response to be analyzed on the same link.
    \item We identify how spherical-wave-induced beam-projection reordering interacts with multilayer codebook constraints. The paired-channel singular-mode gains and SVD-EP rates show no systematic FF/NF displacement, although individual users and modes can strengthen or weaken and the projection gains and ordering of plane-wave candidates change accordingly. The finite codebook then reselects beams within its legal multilayer codeword set. Type-I rank-dependent indices constrain which beams can be combined and may cause the rate-maximizing codeword to omit a candidate with a higher individual projection gain. eType-II retains more angularly separated strong components through multibasis-beam selection and complex-weight adjustment, and attains a higher absolute rate at high ranks in both scenarios.
    \item We quantify the relative benefits of multilayer spatial representation and finite-range sampling. Nested candidate sets show that relaxing the Type-I multi-layer spatial-combination constraint provides the dominant gain, whereas finite range states yield a smaller additional improvement. The eType-II multi-beam extension produces comparable gains over FF and NF channels. These results support prioritizing multilayer multibeam combination and complex-weight representation in future large-port NR CSI codebooks, with range states providing complementary compensation for curvature mismatch.
\end{itemize}

\subsection{Paper Organization}

Section~\ref{sec:method} presents the paired channels, rate metrics, codebook structures, and simulation settings. Section~\ref{sec:response} proceeds from the LoS geometric reference to singular-mode gains, absolute rates of existing codebooks, the response obtained by applying the FF-selected codeword directly to the NF channel, and a representative user's result after reselection. Section~\ref{sec:extensions} compares the rate gains and feedback costs obtained by relaxing Type-I's fixed interlayer offsets, adding finite range states, and increasing the eType-II basis-beam count. Section~\ref{sec:conclusion} concludes the paper.

\section{System Model and Evaluation Methodology}
\label{sec:method}

This section presents the paired FF/NF channels, downlink signal model, evaluation metrics, codebook structures, and simulation settings. Every FF/NF difference reported below is obtained for the same user, the same stochastic multipath parameters, and the same link budget.

\subsection{Paired FF/NF 3GPP Channel Model}

\begin{figure*}[!t]
    \centering
    \includegraphics[width=0.7\textwidth]{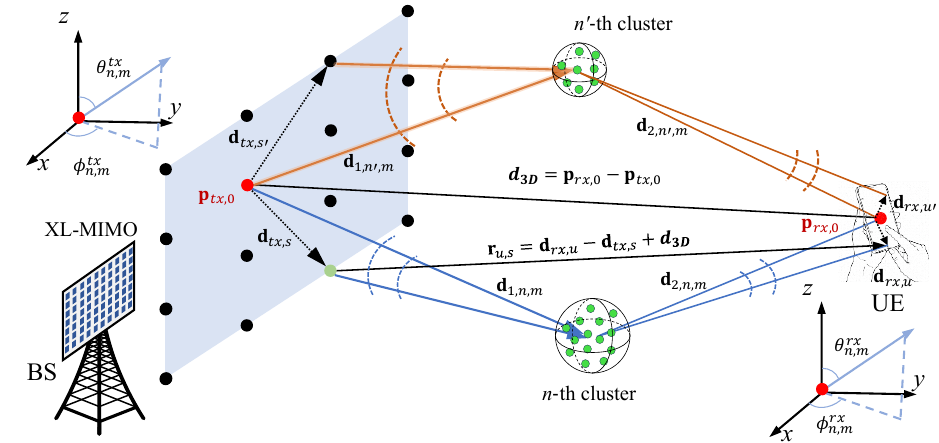}
    \caption{Element-level geometry of the TR~38.901 near-field channel model.}
    \label{fig:system}
\end{figure*}

We consider a single-cell downlink in which a BS with an \(N_{\mathrm t}\)-port dual-polarized XL array serves a UE with \(N_{\mathrm r}=4\) receive ports. The transmitter selects a rank-\(\nu\) precoding matrix from a finite codebook, and the receiver has ideal instantaneous CSI. Fig.~\ref{fig:system} shows the element-level geometry of the TR~38.901 near-field channel, including the transmit and receive array reference points, element offsets, the LoS element-pair distance, and the transmit and receive distances and angles of rays within NLoS clusters.

On the geometry in Fig.~\ref{fig:system}, we apply the element-level near-field extension of TR~38.901 Release~19 and construct strictly paired FF/NF channels~\cite{3gpp38901Rel19}. The two branches share the user location, cluster and ray parameters, polarization, path loss, and noise conditions, and differ only in the element-level wavefront model. Let \(s=1,\ldots,N_{\mathrm t}\) and \(u=1,\ldots,N_{\mathrm r}\) index the transmit and receive ports, respectively, and let \(X\in\{\mathrm{FF},\mathrm{NF}\}\) denote the wavefront model. The small-scale channel impulse response is
\begin{equation}
\begin{aligned}
H^{(X)}_{u,s}(\tau,t)
:={}&\sqrt{\frac{1}{K_{\mathrm R}+1}}H^{\mathrm{NLoS},(X)}_{u,s}(\tau,t)\\
&+\sqrt{\frac{K_{\mathrm R}}{K_{\mathrm R}+1}}h^{\mathrm{LoS},(X)}_{u,s}(t)\delta(\tau-\tau_{\mathrm{LoS}}),
\end{aligned}
\label{eq:rician-cir}
\end{equation}
where \(K_{\mathrm R}\) is the linear Rician \(K\)-factor. In Fig.~\ref{fig:system}, \(\mathbf p_{\mathrm{tx},0}\) and \(\mathbf p_{\mathrm{rx},0}\) are the transmit- and receive-array reference points, and \(\mathbf d_{\mathrm{tx},s}\) and \(\mathbf d_{\mathrm{rx},u}\) are element-offset vectors relative to those points. Bold symbols denote displacement vectors, and the corresponding nonbold symbols denote their Euclidean norms. The reference-point displacement and the displacement of transmit--receive element pair \((u,s)\) are
\begin{equation}
\begin{aligned}
\mathbf d_{3\mathrm D}
&=\mathbf p_{\mathrm{rx},0}-\mathbf p_{\mathrm{tx},0},\\
d_{3\mathrm D}&=\|\mathbf d_{3\mathrm D}\|,\\
\mathbf r_{u,s}
&=\mathbf d_{\mathrm{rx},u}-\mathbf d_{\mathrm{tx},s}
+\mathbf d_{3\mathrm D},\\
r_{u,s}&=\|\mathbf r_{u,s}\|.
\end{aligned}
\label{eq:geometry-vectors}
\end{equation}
For the NF branch, the LoS coefficient uses the exact element-pair distance \(r_{u,s}\):
\begin{equation}
\begin{aligned}
h^{\mathrm{LoS},(\mathrm{NF})}_{u,s}(t)
:={}&\mathbf F_{\mathrm{rx},u}^{\mathrm T}
\!\left(\theta_{\mathrm{LoS}}^{\mathrm{rx}},
\phi_{\mathrm{LoS}}^{\mathrm{rx}}\right)\\
&\times\boldsymbol\Gamma_{\mathrm{LoS}}
\mathbf F_{\mathrm{tx},s}
\!\left(\theta_{\mathrm{LoS}}^{\mathrm{tx}},\phi_{\mathrm{LoS}}^{\mathrm{tx}}\right)\\
&\times\exp\!\left(-j\frac{2\pi}{\lambda}
    r_{u,s}\right)\\
&\times\exp\!\left(j2\pi\nu_{\mathrm{LoS}}t\right).
\end{aligned}
\label{eq:los-coefficient}
\end{equation}
The dual-polarized element patterns are denoted by \(\mathbf F_{\mathrm{tx},s}(\theta,\phi)\) and \(\mathbf F_{\mathrm{rx},u}(\theta,\phi)\), \(\boldsymbol\Gamma_{\mathrm{LoS}}\) is the LoS polarization-coupling matrix, and \(\nu_{\mathrm{LoS}}\) is the Doppler shift. The FF branch retains the same path angles, element patterns, polarization matrix, and Doppler term but replaces the exact spherical-wave phase specified by \(r_{u,s}\) in \eqref{eq:los-coefficient} with the first-order plane-wave phase at the array reference points.

The NLoS channel is the superposition of multiple clusters and rays:
\begin{equation}
H^{\mathrm{NLoS},(X)}_{u,s}(\tau,t)
:=\sum_{n=1}^{N_{\mathrm{cl}}}\sum_{m=1}^{M}h^{(X)}_{u,s,n,m}(t)\delta(\tau-\tau_{n,m}),
\label{eq:nlos-cir}
\end{equation}
where
\begin{equation}
\begin{aligned}
h^{(X)}_{u,s,n,m}(t)
:={}&\sqrt{\frac{P_n}{M}}\,
\mathbf F_{\mathrm{rx},u}^{\mathrm T}
\!\left(\theta^{\mathrm{rx}}_{n,m},\phi^{\mathrm{rx}}_{n,m}\right)\\
&\times\boldsymbol\Gamma_{n,m}
\mathbf F_{\mathrm{tx},s}
\!\left(\theta^{\mathrm{tx}}_{n,m},\phi^{\mathrm{tx}}_{n,m}\right)\\
&\times\exp\!\left(j\psi^{(X)}_{u,s,n,m}\right)
\exp\!\left(j2\pi\nu_{n,m}t\right).
\end{aligned}
\label{eq:nlos-ray-coefficient}
\end{equation}
Here, \(N_{\mathrm{cl}}\) is the number of NLoS clusters, \(M\) is the number of rays per cluster, \(P_n\) is the power of cluster \(n\), and \(\tau_{n,m}\), \(\nu_{n,m}\), and \(\boldsymbol\Gamma_{n,m}\) denote the ray delay, Doppler shift, and polarization-coupling matrix, respectively. Physical zenith and azimuth angles are denoted generically by \((\theta,\phi)\). Let \(\mathbf d_{1,n,m}\) and \(\mathbf d_{2,n,m}\) be the displacement vectors from the transmit and receive reference points to the corresponding spherical-wave sources of ray \((n,m)\), and let \(d_{q,n,m}=\|\mathbf d_{q,n,m}\|\), \(q\in\{1,2\}\). After removing the common phase at the array reference points, the NF and FF element-level phases are~\cite{xuhuixin-3GPP-channel}
\begin{equation}
\begin{aligned}
\psi^{\mathrm{NF}}_{u,s,n,m}
=\frac{2\pi}{\lambda}\Big[&d_{1,n,m}
  -\|\mathbf d_{1,n,m}-\mathbf d_{\mathrm{tx},s}\|\\
&+d_{2,n,m}
  -\|\mathbf d_{2,n,m}-\mathbf d_{\mathrm{rx},u}\|\Big],\\
\psi^{\mathrm{FF}}_{u,s,n,m}
=\frac{2\pi}{\lambda}\Big(&
  \frac{\mathbf d_{1,n,m}^{\mathrm T}\mathbf d_{\mathrm{tx},s}}
  {d_{1,n,m}}
  +\frac{\mathbf d_{2,n,m}^{\mathrm T}\mathbf d_{\mathrm{rx},u}}
  {d_{2,n,m}}\Big).
\end{aligned}
\label{eq:paired-path-phases}
\end{equation}
The directions of \(\mathbf d_{1,n,m}\) and \(\mathbf d_{2,n,m}\) are determined by the transmit- and receive-side zenith and azimuth angles shown in Fig.~\ref{fig:system}. As \(d_{1,n,m}\) and \(d_{2,n,m}\) increase, the first-order expansion of the exact distances makes \(\psi^{\mathrm{NF}}_{u,s,n,m}\) converge to \(\psi^{\mathrm{FF}}_{u,s,n,m}\). The same ray therefore differs only in whether it uses spherical distances or the linear phase determined by the reference-point directions.

At the selected time \(t_0\), the LoS component and all NLoS ray coefficients are summed directly to form the single-frequency narrowband spatial channel used in the subsequent evaluation:
\begin{equation}
\begin{aligned}
\left[\widetilde{\mathbf H}_{X}(t_0)\right]_{u,s}
={}&\sqrt{\frac{1}{K_{\mathrm R}+1}}
\sum_{n=1}^{N_{\mathrm{cl}}}\sum_{m=1}^{M}h^{(X)}_{u,s,n,m}(t_0)\\
&+\sqrt{\frac{K_{\mathrm R}}{K_{\mathrm R}+1}}
h^{\mathrm{LoS},(X)}_{u,s}(t_0).
\end{aligned}
\label{eq:single-frequency-channel}
\end{equation}
Let \(g_{\mathrm{LS}}\) be the linear large-scale power gain due to path loss and shadow fading. The rate calculation uses \(\mathbf H_X=\sqrt{g_{\mathrm{LS}}}\widetilde{\mathbf H}_{X}(t_0)\). For each user drop, the location, large-scale parameters, cluster powers, delays, ray angles, polarization matrices, Doppler shifts, path loss, and shadow fading are generated once and shared by the FF and NF branches. The two branches use only the plane-wave and spherical-wave element phases defined above. Spatially non-stationary power variation is disabled, and the complex amplitude of every ray is retained, thereby restricting each paired difference to the spatial-channel and codebook-matching changes induced by spherical-wave curvature.

\subsection{Rate Model and Evaluation Benchmark}

Let \(\mathbf s\in\mathbb C^{\nu}\) denote the transmit symbol vector with \(\mathbb E[\mathbf s\mathbf s^{\mathrm H}]=\mathbf I_{\nu}\), and let \(P_{\mathrm t}\) be the total transmit power. For a rank-\(\nu\) precoder \(\mathbf W\in\mathbb C^{N_{\mathrm t}\times\nu}\), equal power is assigned to all layers and \(\|\mathbf W\|_{\mathrm F}^{2}=1\). The received signal is \(\mathbf y_X=\sqrt{P_{\mathrm t}}\mathbf H_X\mathbf W\mathbf s+\mathbf n\), where \(\mathbf n\sim\mathcal{CN}(\mathbf 0,\sigma_n^2\mathbf I_{N_{\mathrm r}})\). The noise power \(\sigma_n^2=N_0BF_{\mathrm N}\) is determined by the noise power spectral density \(N_0\), bandwidth \(B\), and linear receiver noise figure \(F_{\mathrm N}\). With \(\rho=P_{\mathrm t}/\sigma_n^2\), the achievable rate is~\cite{telatar1999capacity}
\begin{equation}
R_\nu(\mathbf H,\mathbf W)
=\log_2\det\!\left(
\mathbf I+\rho\mathbf H\mathbf W\mathbf W^{\mathrm H}\mathbf H^{\mathrm H}
\right),
\label{eq:achievable-rate}
\end{equation}
Path loss and shadow fading have already been applied to \(\mathbf H_X\) through \(g_{\mathrm{LS}}\) and are not included again in \(\rho\). Let \(\mathcal C^{(c,\nu)}\) denote the rank-\(\nu\) candidate set of codebook \(c\). The receiver selects a codeword from instantaneous CSI and feeds back its index~\cite{love2005limited}; the optimal codeword is
\begin{equation}
\mathbf W^{(c,\nu),\star}(\mathbf H)
=\arg\max_{\mathbf W\in\mathcal C^{(c,\nu)}}
R_\nu(\mathbf H,\mathbf W).
\label{eq:codeword-selection}
\end{equation}
We use rank-\(\nu\) SVD precoding with equal power allocation (SVD-EP) as the benchmark. For \(\mathbf H=\mathbf U\boldsymbol\Sigma\mathbf V^{\mathrm H}\), let \(\mathbf v_i\) and \(\sigma_i\) denote the \(i\)th right singular vector and singular value, respectively. Then
\begin{equation}
\begin{aligned}
\mathbf W_{\mathrm{SVD\text{-}EP}}^{(\nu)}
&=\frac{1}{\sqrt{\nu}}[\mathbf v_1,\ldots,\mathbf v_\nu],\\
R_{\mathrm{SVD\text{-}EP}}^{(\nu)}(\mathbf H)
&=\sum_{i=1}^{\nu}\log_2\!\left(1+\frac{\rho}{\nu}\sigma_i^2\right),
\end{aligned}
\label{eq:svd-ep-rate}
\end{equation}
SVD-EP divides the total power equally among the first \(\nu\) channel eigenmodes. Type-I, eType-II, and SVD-EP are compared at the same rank, total transmit power, and equal layer-power constraint.

\subsection{Type-I/eType-II Codebooks and Port Extrapolation}

The Type-I single-panel codebook uses oversampled two-dimensional DFT beams. Let the single-polarization port array be \(N_1\times N_2\), with oversampling factors \(O_1\) and \(O_2\), so that \(P_{\mathrm{CSI-RS}}=2N_1N_2\). A spatial basis beam is \(\mathbf v_{l,m}=\mathbf q_l\otimes\mathbf u_m\), where
\begin{subequations}
\begin{align}
\mathbf q_l
&=\left[1,e^{j2\pi l/(O_1N_1)},\ldots,
e^{j2\pi l(N_1-1)/(O_1N_1)}\right]^{\mathrm T},\\
\mathbf u_m
&=\left[1,e^{j2\pi m/(O_2N_2)},\ldots,
e^{j2\pi m(N_2-1)/(O_2N_2)}\right]^{\mathrm T}.
\end{align}
\label{eq:dft-bases}
\end{subequations}
Let \(\gamma_n=e^{j\pi n/2}\) denote the quantized polarization co-phase. A rank-1 codeword is
\begin{equation}
\mathbf w^{(1)}_{l,m,n}
=\frac{1}{\sqrt{P_{\mathrm{CSI-RS}}}}
\begin{bmatrix}
\mathbf v_{l,m}\\
\gamma_n\mathbf v_{l,m}
\end{bmatrix},
\label{eq:type1-r1}
\end{equation}
and a rank-2 codeword can be represented as
\begin{equation}
\mathbf W^{(2)}_{l,l',m,m',n}
=\frac{1}{\sqrt{2P_{\mathrm{CSI-RS}}}}
\begin{bmatrix}
\mathbf v_{l,m}&\mathbf v_{l',m'}\\
\gamma_n\mathbf v_{l,m}&-\gamma_n\mathbf v_{l',m'}
\end{bmatrix}.
\label{eq:type1-r2}
\end{equation}
The second basis beam \((l',m')\) is not selected independently; PMI index \(i_{1,3}\) maps to a finite rank-dependent offset \((k_1,k_2)\), and ranks~3--4 use the corresponding legal combinations~\cite{3gpp38214Rel19}. Type-I therefore has low feedback overhead, while the beams used by different layers are coupled through common indices and predefined offsets, which limits the simultaneous coverage of angularly separated multipath clusters.

The frequency-flat eType-II codebook selects \(L\) basis beams from one common oversampled subgrid. With \(\mathbf B=[\mathbf v_{l_1,m_1},\ldots,\mathbf v_{l_L,m_L}]\), its spatial precoder can be written as
\begin{equation}
\mathbf W=
\begin{bmatrix}
\mathbf B&\mathbf 0\\
\mathbf 0&\mathbf B
\end{bmatrix}
[\mathbf c_1,\ldots,\mathbf c_\nu],
\label{eq:etype}
\end{equation}
where \(\mathbf c_r\in\mathbb C^{2L}\) contains the sparse amplitude and phase coefficients of layer \(r\). The complete oversampled grid comprises \(O_1O_2\) DFT subgrids. An eType-II codeword first selects a common offset \((q_1,q_2)\), then chooses an unordered set of \(L\) beams from the corresponding subgrid and quantizes the nonzero-coefficient locations, amplitudes, and phases for every polarization and layer. We retain the common offset, basis-beam set, nonzero-coefficient constraints, and rank-dependent rules. We set \(N_3=M_\nu=1\) to exclude cross-subband compression from the spatial comparison. The baseline eType-II codebook uses \(L=4\); Section~\ref{sec:extensions} extends the same structure to \(L=6\) and \(L=8\).

The 160- and 4418-port configurations exceed the port range specified in TS~38.214. Each dual-polarized element is mapped to an independent port, and the Type-I/eType-II constructions above are evaluated in the extrapolated dimensions. The extrapolation retains the codebook structures that determine multi-layer spatial representation. The terms Type-I and eType-II below follow this convention.

\begin{table*}[!t]
\caption{Paired FF/NF Link and Codebook Simulation Configuration}
\label{tab:configuration}
\centering
\footnotesize
\setlength{\tabcolsep}{5pt}
\begin{tabular}{p{0.20\textwidth}p{0.36\textwidth}p{0.36\textwidth}}
\toprule
\textbf{Parameter} & \textbf{UMi} & \textbf{InH-linear}\\
\midrule
Carrier frequency & 7~GHz & 24~GHz\\
BS/UT height & 10/1.5~m & 3/1~m\\
User region & \(d_{2\mathrm D}\in[10,115.5]\)~m & \(d_{2\mathrm D}\leq14.14\)~m\\
BS array per polarization & \(47\times47\), half-wavelength spacing & \(80\times1\), half-wavelength spacing\\
Port mapping and total ports & One-to-one, 4418 & One-to-one, 160\\
Aperture \(D\) / Rayleigh distance & 1.393/90.6~m & 0.494/39.0~m\\
UT array & \(2\times1\) dual-polarized, four ports & \(2\times1\) dual-polarized, four ports\\
Total transmit power / link-budget bandwidth & 24~dBm / 100~MHz & 24~dBm / 400~MHz\\
Noise PSD / noise figure & \(-174\)~dBm/Hz / 7~dB & \(-174\)~dBm/Hz / 7~dB\\
Channel-frequency sample & Single point at carrier & Single point at carrier\\
Evaluated rank & 1--4 & 1--4\\
\bottomrule
\end{tabular}
\end{table*}

\subsection{Simulation Configuration}

Table~\ref{tab:configuration} summarizes the two principal settings. Let \(d_{2\mathrm D}\) be the horizontal-plane projection distance between the BS and UT array reference points, with endpoint heights \(h_{\mathrm{BS}}\) and \(h_{\mathrm{UT}}\). The three-dimensional distance in Fig.~\ref{fig:system} is then \(d_{3\mathrm D}=\sqrt{d_{2\mathrm D}^2+(h_{\mathrm{BS}}-h_{\mathrm{UT}})^2}\). UMi uses the \(47\times47\)-element-per-polarization planar array permitted by the TR~38.901 near-field calibration, and InH-linear uses an 80-element-per-polarization ULA. The aperture \(D\) in the Rayleigh distance is the largest separation between two elements of one polarization~\cite{selvan2017fraunhofer}: the planar diagonal \(D=1.393\)~m for UMi and the end-to-end span \(D=0.494\)~m for InH-linear. The corresponding \(d_{\mathrm R}=2D^2/\lambda\) values are 90.6 and 39.0~m.

Each user has one narrowband spatial channel matrix generated at the carrier frequency, while the receiver noise in the link budget is calculated over the 100- and 400-MHz system bandwidths listed in Table~\ref{tab:configuration}. The UE has ideal instantaneous CSI, and inter-cell interference and channel coding are excluded. Fixed-rank evaluation is used to isolate the codebook's multi-layer representation capability.

\section{Near-Field Channel Characteristics and Codebook Response}
\label{sec:response}

This section examines how spherical-wave propagation changes the multi-layer response of existing plane-wave codebooks. The LoS geometry provides separate references for the modal scale of double-sided arrays and the single-path coherent mismatch of a fixed plane-wave codeword. Paired UMi and InH-linear channels then reveal the corresponding changes in singular modes, SVD-EP benchmark rates, and multi-layer Type-I/eType-II rates.

\subsection{Theoretical Analysis of Spherical-Wave Effects on the LoS Path}

For a near-field LoS MIMO link formed by two non-parallel ULAs, a prolate-spheroidal-wave-function-based approximation characterizes the spatial DoF scaling as~\cite{wu2022distanceaware}
\begin{equation}
\begin{aligned}
N_{\mathrm{DoF}}(r)
&\approx
\frac{(N_{\mathrm t}-1)(N_{\mathrm r}-1)d^2}{\lambda r}
|\cos\theta\cos\phi|\\
&=\frac{D_{\mathrm{t,proj}}D_{\mathrm{r,proj}}}{\lambda r}.
\end{aligned}
\label{eq:los-coupling}
\end{equation}
where \(d\) is the element spacing, \(r\) is the link distance, \(\theta\) and \(\phi\) specify the array orientations, and \(D_{\mathrm{t,proj}}\) and \(D_{\mathrm{r,proj}}\) are the corresponding projected apertures. The two-position UE in both scenarios has \(D_{\mathrm{UE}}=\lambda/2\). Setting both projection factors to unity, using the UMi planar diagonal or the InH-linear end-to-end span as the maximum one-dimensional BS aperture, and substituting the shortest sampled 3D distance yields maximum values of 0.0433 and 0.1148 for the approximation in~\eqref{eq:los-coupling}, respectively.
The far-field LoS channel has one spatial DoF~\cite{wu2022distanceaware}. The approximation in~\eqref{eq:los-coupling} characterizes the growth of LoS spatial DoF with aperture coupling and should not be interpreted as a literal mode count when its value falls below unity. The values obtained here are far below unity, suggesting that additional appreciable LoS spatial modes are unlikely in the configurations considered.

The second scale is the coherent mismatch between a plane-wave codeword and one spherical-wave path. Consider an \(N\)-element ULA with spacing \(d=\lambda/2\) and its origin at the array center. Element \(n\) is located at \(x_n=[n-(N+1)/2]d\), where \(n=1,\ldots,N\), and the aperture is \(D=(N-1)d\). Let a spherical-wave source be located at center range \(r\) with array-axis direction cosine \(\Omega\). Its exact distance to element \(n\) is \(r_n=\sqrt{r^2+x_n^2-2rx_n\Omega}\).

With \(k=2\pi/\lambda\), we neglect the slowly varying amplitude across the aperture and remove the common phase associated with \(r\). For a plane-wave codeword pointing toward \(\Omega_{\rm c}=\Omega+\Delta\Omega\), the unit-norm FF codeword and NF response are \([\mathbf a(\Omega_{\rm c})]_n=N^{-1/2}e^{jkx_n\Omega_{\rm c}}\) and \([\mathbf b(\Omega,r)]_n=N^{-1/2}e^{-jk(r_n-r)}\), respectively. Their exact normalized coherent gain is \(G(\Delta\Omega,r)=|\mathbf a^{\mathrm H}(\Omega+\Delta\Omega)\mathbf b(\Omega,r)|^2\). In the Fresnel region, the exact path length admits the second-order expansion \(r_n-r\simeq-x_n\Omega+x_n^2(1-\Omega^2)/(2r)\)~\cite{cui2022polar}.
The first-order term \(-x_n\Omega\) cancels the plane-wave phase aligned with the physical path. The remaining linear phase is determined by \(\Delta\Omega\), whereas the quadratic term retains the wavefront curvature. With \(d_{\mathrm R}=2D^2/\lambda\) and normalized range \(\alpha=r/d_{\mathrm R}\), substitution gives
\begin{equation}
G(\Delta\Omega,\alpha)
\simeq
\left|\frac{1}{N}\sum_{n=1}^{N}
e^{-j2\pi x_n\Delta\Omega/\lambda}
e^{-j\frac{\pi}{2\alpha}(x_n/D)^2(1-\Omega^2)}
\right|^2.
\label{eq:joint-mismatch}
\end{equation}
The two exponential factors capture the linear phase from angular error and the quadratic phase from wavefront curvature, respectively. In the far-field limit, the quadratic phase vanishes and the half-wavelength ULA response reduces to the standard DFT-array response.

For \(\Delta\Omega=0\), the FF baseline is unity and the finite-range reduction is entirely due to the quadratic phase. For \(\Delta\Omega\neq0\), the FF baseline is already reduced by the DFT response, while the linear and quadratic phases remain coupled within the same coherent sum and their losses are not separable. Fig.~\ref{fig:los-mismatch} uses \(N=80\), \(\Omega=0\), and normalizes \(\Delta\Omega\) by the base DFT spacing \(\Delta\Omega_{\mathrm{DFT}}=2/N\). The maximum offset to the nearest DFT codeword is one half-grid spacing, so the curves cover \(0\leq\Delta\Omega/\Delta\Omega_{\mathrm{DFT}}\leq1/2\). As range decreases, curvature produces the largest reduction for the otherwise aligned codeword; with an existing angular offset, the additional finite-range reduction is generally smaller. The convergence and crossing of the curves in the deeper near field follow directly from the joint linear--quadratic phase accumulation.

\addtocounter{figure}{1}
\begin{figure*}[!t]
    \centering
    \includegraphics[width=\textwidth]{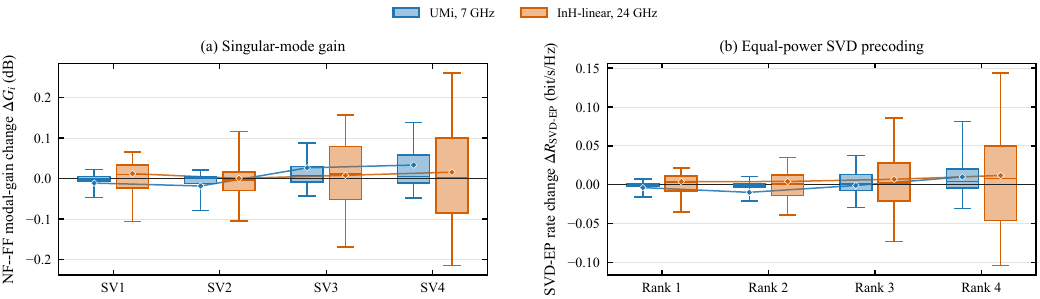}
    \caption{Paired FF-to-NF changes of (a) singular-mode gains and (b) rank-matched SVD-EP rates for UMi at 7~GHz and InH-linear at 24~GHz. Boxes, whiskers, center lines, and diamonds denote P25--P75, P10--P90, medians, and means.}
    \label{fig:modal-svd}
\end{figure*}
\addtocounter{figure}{-2}

These LoS results motivate two paired-channel tests: whether the FF/NF switch preserves the strong modes and SVD-EP rate, and how the distance-dependent mismatch of a fixed codeword appears in the final Type-I and eType-II rates.

\begin{figure}[!h]
    \centering
    \includegraphics[width=\columnwidth]{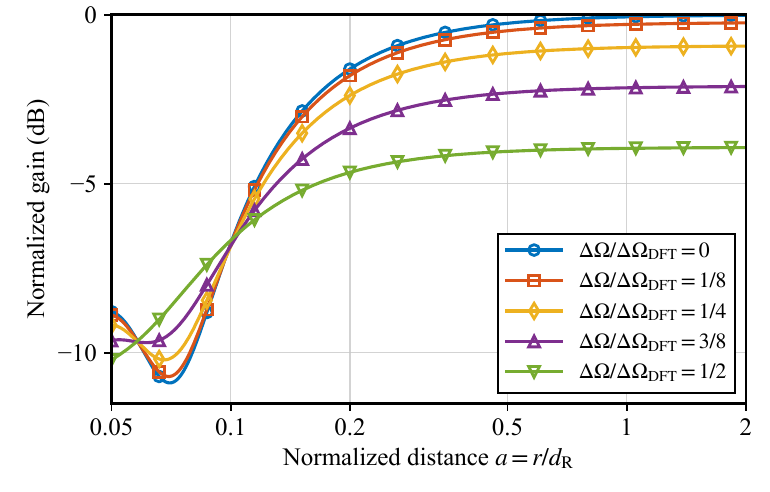}
    \caption{Joint distance- and angular-mismatch gain of a far-field DFT codeword against a near-field LoS response for an \(N=80\) half-wavelength ULA.}
    \label{fig:los-mismatch}
\end{figure}
\addtocounter{figure}{1}

\subsection{Paired-Channel Singular Modes and SVD-EP Rates}

For \(X\in\{\mathrm{FF},\mathrm{NF}\}\), define the paired change of singular mode \(i\) and the rank-\(\nu\) SVD-EP rate change as
\begin{equation}
\begin{aligned}
\Delta G_i
&=20\log_{10}\frac{\sigma_{i,\mathrm{NF}}}{\sigma_{i,\mathrm{FF}}},\\
\Delta R_{\mathrm{SVD\text{-}EP}}^{(\nu)}
&=R_{\mathrm{SVD\text{-}EP,NF}}^{(\nu)}
-R_{\mathrm{SVD\text{-}EP,FF}}^{(\nu)}.
\end{aligned}
\label{eq:paired-modal-svd-change}
\end{equation}

Fig.~\ref{fig:modal-svd} summarizes paired results for area-uniform users in each scenario. The median \(\Delta G_i\) values lie between \(-0.0003\) and \(0.0075\)~dB in UMi and between \(-0.0034\) and \(0.0120\)~dB in InH-linear. The user spread increases for weaker modes: the SV4 P10--P90 interval is \([-0.0479,0.138]\)~dB in UMi and \([-0.215,0.261]\)~dB in InH-linear, while both distributions remain centered near zero.

The SVD-EP rates follow the same trend. At rank~4, the P10--P90 intervals of \(\Delta R_{\mathrm{SVD\text{-}EP}}^{(\nu)}\) are \([-0.0309,0.0814]\) and \([-0.104,0.144]\)~bit/s/Hz in UMi and InH-linear, respectively; lower ranks have narrower intervals. The near-field array responses in these configurations produce no discernible systematic SVD-EP rate gain. Changes in the higher-order singular modes and SVD-EP rate still cross zero, so a distribution centered near zero does not imply that every user's modal gains remain unchanged. The finite-codebook results below examine how these user-dependent changes affect plane-wave projections and codeword selection.

\subsection{Absolute Rate and SVD-EP-Normalized Rate}

\begin{figure*}[!t]
    \centering
    \includegraphics[width=\textwidth]{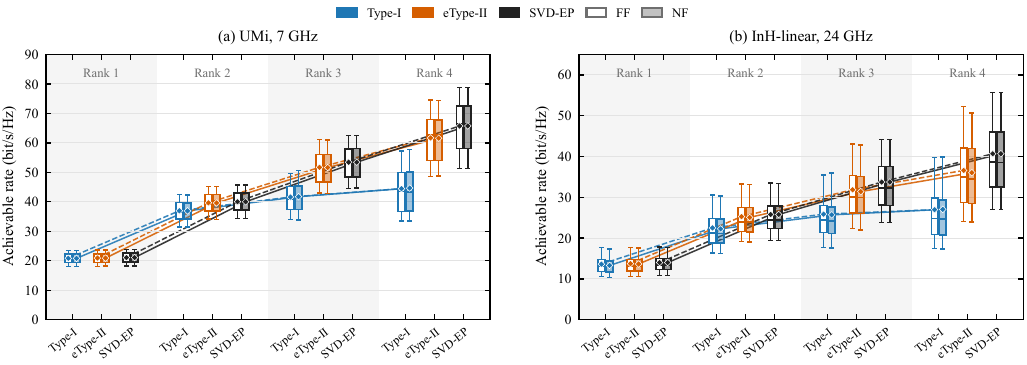}
    \caption{Achievable-rate distributions of Type-I, eType-II \(L=4\), and rank-matched SVD-EP under paired FF/NF channels in UMi at 7~GHz and InH-linear at 24~GHz.}
    \label{fig:absolute-rate}
\end{figure*}

\begin{figure*}[!t]
    \centering
    \includegraphics[width=\textwidth]{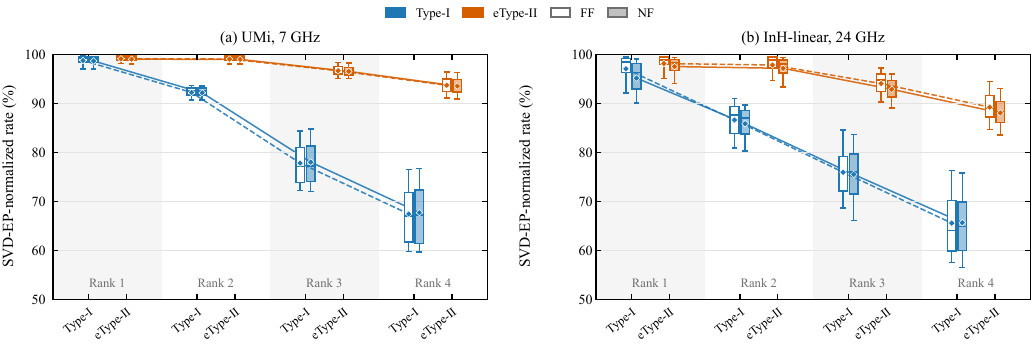}
    \caption{Rates of Type-I and eType-II \(L=4\) normalized by the rank-matched SVD-EP benchmark in UMi at 7~GHz and InH-linear at 24~GHz. Empty and filled boxes denote FF and NF, respectively.}
    \label{fig:svd-rate}
\end{figure*}

Fig.~\ref{fig:absolute-rate} compares the rank-1--4 FF/NF absolute rates of Type-I, eType-II with \(L=4\), and SVD-EP, ordered from left to right within each rank. Throughout the paper, empty and filled boxes denote paired FF and NF results; unless stated otherwise, boxes, whiskers, center lines, and diamonds denote P25--P75, P10--P90, medians, and means.

At rank~1, both finite codebooks can follow one dominant direction and achieve similar absolute rates. As rank increases, the constrained Type-I layer combinations create an increasing separation from eType-II and SVD-EP. At rank~4, the mean NF rates of Type-I and eType-II are 44.7 and 61.6~bit/s/Hz in UMi and 27.0 and 36.0~bit/s/Hz in InH-linear, respectively. By comparison, the FF and NF boxes of the same scheme largely overlap, so the aggregate displacement caused by the wavefront model is much smaller than the separation between codebook structures.

The aggregate FF/NF displacement is small, while individual users may still experience appreciable changes. To compare the realized codebook rates across scenarios and ranks, normalize each user's rate by the SVD-EP benchmark on the same channel:
\begin{equation}
\bar R_{c,X}^{(\nu)}
=\frac{R_{c,X}^{(\nu)}}
{R_{\mathrm{SVD\text{-}EP},X}^{(\nu)}},
\qquad X\in\{\mathrm{FF},\mathrm{NF}\}.
\label{eq:svd-normalized-rate}
\end{equation}
The SVD-EP benchmark corresponds to \(\bar R_{c,X}^{(\nu)}=1\). Fig.~\ref{fig:svd-rate} displays this ratio as a percentage.

The paired FF/NF results in Fig.~\ref{fig:svd-rate} first quantify how the wavefront change affects the realized codebook rate. In UMi, the FF/NF boxes nearly overlap for every rank and codebook, and the difference between their mean normalized rates is at most 0.28 percentage points. In InH-linear, most NF boxes shift downward: the mean normalized rate decreases by 1.89 percentage points for Type-I at rank~1 and by 1.19 percentage points for eType-II at rank~4, with smaller changes for the remaining combinations. The deeper InH-linear near field therefore slightly lowers the SVD-EP-normalized rate realized by the far-field codebooks, while this effect is weak in UMi.

The structural separation between codebooks grows rapidly with rank. At rank~4, the mean NF normalized rates of Type-I and eType-II are \(67.7\%\) and \(93.6\%\) in UMi and \(65.7\%\) and \(88.1\%\) in InH-linear, respectively. Thus, eType-II retains approximately 88--94\% of the SVD-EP rate in both scenarios, whereas Type-I retains approximately 66--68\%. The multiple basis beams and complex coefficients of eType-II cover angularly separated multipath clusters and adjust their relative weights, whereas the constrained Type-I multi-layer beam combinations form the principal high-rank bottleneck in approaching SVD-EP.

\subsection{Frozen-Codeword Response and Range Dependence}

The final FF/NF total-rate distributions in Fig.~\ref{fig:absolute-rate} are close, whereas Fig.~\ref{fig:los-mismatch} shows that a fixed plane-wave codeword can incur appreciable loss in the deep near field. We therefore compare the fixed-codeword mismatch with the final link result. For any codebook \(c\), define
\begin{subequations}
\begin{align}
R_{\mathrm{FF}}
&=R_\nu\!\left(\mathbf H_{\mathrm{FF}},
\mathbf W^\star(\mathbf H_{\mathrm{FF}})\right),\\
R_{\mathrm{NF}}^{\mathrm{frz}}
&=R_\nu\!\left(\mathbf H_{\mathrm{NF}},
\mathbf W^\star(\mathbf H_{\mathrm{FF}})\right),\\
R_{\mathrm{NF}}
&=R_\nu\!\left(\mathbf H_{\mathrm{NF}},
\mathbf W^\star(\mathbf H_{\mathrm{NF}})\right).
\end{align}
\label{eq:three-rates}
\end{subequations}
Their changes relative to the same user's FF baseline are
\begin{equation}
\delta_{\mathrm{frz}}
=\frac{R_{\mathrm{NF}}^{\mathrm{frz}}-R_{\mathrm{FF}}}{R_{\mathrm{FF}}},
\qquad
\delta_{\mathrm{tot}}
=\frac{R_{\mathrm{NF}}-R_{\mathrm{FF}}}{R_{\mathrm{FF}}}.
\label{eq:rate-decomp}
\end{equation}

\begin{figure*}[!t]
    \centering
    \includegraphics[width=\textwidth]{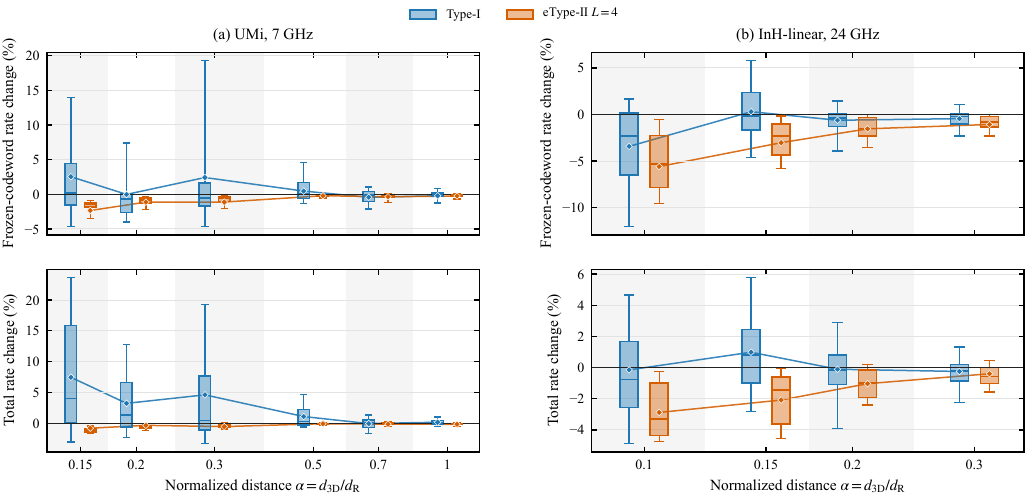}
    \caption{Distance-dependent comparison in UMi at 7~GHz and InH-linear at 24~GHz for Type-I and eType-II \(L=4\) at rank~4. The upper row shows the rate change obtained by freezing the FF-selected codeword; the lower row shows the final FF-to-NF total-rate change.}
    \label{fig:distance-decomp}
\end{figure*}

Both metrics use the same user's FF rate as the baseline, and a negative value denotes a loss relative to FF. The two columns of Fig.~\ref{fig:distance-decomp} correspond to UMi and InH-linear. The upper row gives the rate change after freezing the FF-selected codeword, and the lower row gives the final total-rate change when the codebook operates normally.

Fig.~\ref{fig:distance-decomp} shows a common range dependence. As \(\alpha\) increases, the absolute magnitudes of both the frozen-codeword change and the final total-rate change generally decrease, and the distributions on either side of zero contract toward the FF baseline. The perturbation therefore weakens with range, while its sign depends on the user channel and codebook structure; gains and losses may coexist within the same range group.

The two codebooks exhibit different sign patterns. eType-II is predominantly lossy in both scenarios and across the sampled ranges, with the loss decreasing as range increases. The Type-I distributions straddle zero: some users lose rate, whereas others gain rate. Judged only by their evolution with range, eType-II's frozen-codeword losses are closer to the single-path plane-wave-codeword result in Fig.~\ref{fig:los-mismatch}: finite range introduces a loss that gradually vanishes away from the array. At the same \(\alpha\), eType-II also exhibits similar frozen-codeword losses in the two scenarios.

The Type-I frozen-codeword changes straddle zero because the spherical-wave phase can either weaken or strengthen the multipath coherent response under a prescribed codeword; this occurs before reselection and reflects the joint action of the changed multipath channel and the prescribed codeword. After NF-based reselection, Type-I's legal interlayer combinations further determine how much of the reordered projection gain becomes a final-rate improvement. For the InH-linear \(\alpha=0.10\) group, the median frozen-codeword changes of Type-I and eType-II are \(-2.28\%\) and \(-5.34\%\); after reselection, the corresponding final changes shrink to \(-0.770\%\) and \(-3.28\%\). Positive Type-I changes are more pronounced in UMi. At \(\alpha=0.15\), the median frozen-codeword change is \(0.157\%\), while the median final total-rate change reaches \(4.03\%\), with a P25--P75 interval of \([0.0910\%,15.8\%]\). UMi uses a \(47\times47\) array per polarization, substantially more elements than the \(80\times1\) InH-linear array. On the finer two-dimensional DFT grid, the predefined rank-dependent Type-I offsets occupy a more local portion of the full angular-grid range, so high-rank FF codewords are more often constrained by the allowed combinations. When the NF wavefront changes the path projections on this grid, the energy of some users moves toward a legal combination and produces the more prominent positive tail.

The next subsection tests the beam-reselection mechanism in Fig.~\ref{fig:beam-reordering} through a representative user, examining how the NF wavefront changes the relative strengths of the beams associated with the LoS ray and NLoS multipath clusters and how Type-I and eType-II recombine them within their legal codeword sets.

\subsection{Representative-User Beam Reselection}

\begin{figure*}[!t]
    \centering
    \includegraphics[width=\textwidth]{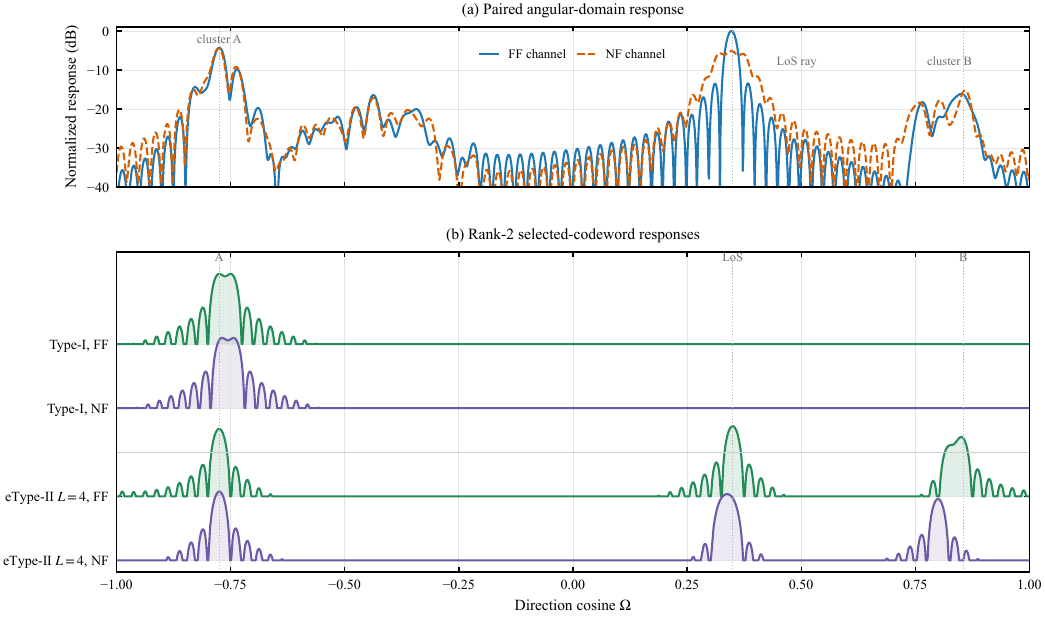}
    \caption{Representative InH-linear paired user: (a) FF/NF angular-domain channel response of one fixed co-polarized branch and (b) ridgeline responses of the rank-2 codewords selected by Type-I and eType-II \(L=4\) on the FF and NF channels.}
    \label{fig:rematching}
\end{figure*}

We select one paired InH-linear user at \(\alpha=0.10\) and examine the mechanism through its multipath response, selected beams, and complete-codebook rates. Fig.~\ref{fig:rematching}(a) projects one fixed co-polarized FF/NF branch onto the same plane-wave dictionary, while Fig.~\ref{fig:rematching}(b) compares the rank-2 responses selected by Type-I and eType-II \(L=4\); the ridgelines are clipped uniformly to \(-25\)--0~dB for comparison. Table~\ref{tab:representative-user} reports the selected beams and the rates computed from the full dual-polarized channel and quantized codewords.

\begin{table*}[!t]
\caption{Selected Beams and Full Dual-Polarized Rates of the Representative InH-Linear User}
\label{tab:representative-user}
\centering
\scriptsize
\setlength{\tabcolsep}{3.6pt}
\begin{tabular}{@{}lcllccc@{}}
\toprule
Codebook & Rank & FF-selected beams \(\Omega\) & NF-selected beams \(\Omega\) & \(R_{\mathrm{FF}}\) & \(R_{\mathrm{NF}}^{\mathrm{frz}}\) & \(R_{\mathrm{NF}}\)\\
\midrule
Type-I & 1 & \(0.350\) & \(-0.775\) & 18.3 & 16.7 & 16.9\\
Type-I & 2 & \(\{-0.775,-0.750\}\) & \(\{-0.769,-0.744\}\) & 24.6 & 24.8 & 24.8\\
eType-II \(L=4\) & 1 & \(\{-0.800,-0.775,-0.725,0.350\}\) & \(\{-0.775,0.325,0.350,0.375\}\) & 18.8 & 17.3 & 18.6\\
eType-II \(L=4\) & 2 & \(\{0.350,-0.775,0.850,0.825\}\) & \(\{-0.775,0.350,0.325,0.800\}\) & 31.0 & 28.6 & 30.0\\
\bottomrule
\end{tabular}
\end{table*}

The LoS ray of the FF subchannel lies near \(\Omega=0.35\), while a strong multipath cluster~A appears near \(\Omega=-0.78\). Under the NF spherical-wave response, cluster~A becomes the global peak, and the Type-I rank-1 beam in Table~\ref{tab:representative-user} consequently switches from \(0.350\) to \(-0.775\). For rank~2 in Fig.~\ref{fig:rematching}(b), the largest legal nonzero offset spans only three base-DFT positions, whereas the peaks associated with the LoS ray and cluster~A are separated by approximately 35 DFT-grid points. Type-I can therefore construct a two-layer codeword in only one of the two local regions. This legal offset already limits their joint representation in the FF channel. After the NF phase changes the beam responses and inter-layer coupling within the local region, reselecting the two-beam combination yields a small improvement. The frozen-codeword and final rank-2 changes are \(+0.704\%\) and \(+0.914\%\), respectively.

eType-II selects a common set of four basis beams for all layers and uses per-layer quantized complex coefficients to adjust each basis beam's amplitude and phase. It can cover the LoS ray and several angularly separated multipath clusters simultaneously. For this user, the rank-1 change improves from \(-8.30\%\) with the frozen codeword to \(-1.48\%\) after reselection; the corresponding rank-2 change improves from \(-7.72\%\) to \(-3.32\%\). Multibeam reselection thus recovers most of the fixed-codeword loss and gives a rank-2 NF rate of 30.0~bit/s/Hz, above the 24.8~bit/s/Hz of Type-I.

\section{Codebook Design Insights and Feedback Cost}
\label{sec:extensions}

Section~III showed no systematic FF/NF displacement in the singular-mode gains or SVD-EP rates of the paired channels, although individual users' modal gains and plane-wave beam projections still change. At high ranks, the main Type-I/eType-II gap instead comes from the codebook's ability to represent several strong spatial directions jointly. We therefore make two comparisons in the 24-GHz InH-linear configuration. The first keeps the angular grid, polarization structure, and power constraint fixed, relaxes Type-I's fixed interlayer beam offsets, and then adds finite range states. The second changes only the number of eType-II plane-wave basis beams. We finally compare the rate increments from these changes with their uncoded CSI-report payloads.

\begin{figure*}[!t]
    \centering
    \includegraphics[width=\textwidth]{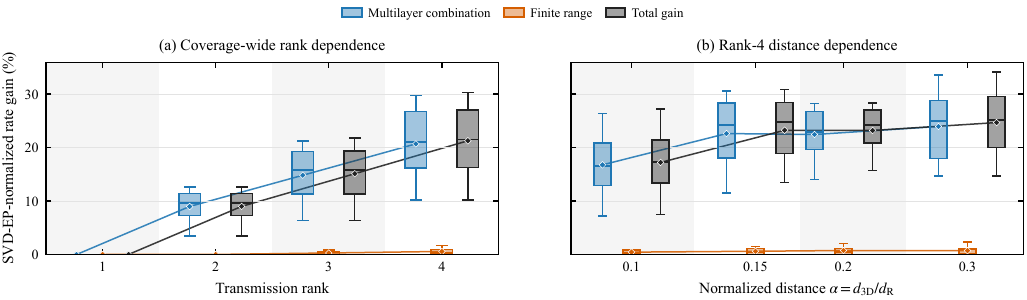}
    \caption{SVD-EP-normalized rate gains from multilayer combination and finite range states in InH-linear at 24~GHz versus rank and versus normalized distance at rank~4.}
    \label{fig:attribution}
\end{figure*}

\subsection{Type-I Interlayer Combination and Finite Range States}

Type-I layers share a base DFT-beam index and use rank-dependent offsets to construct a multilayer codeword. To identify the separate effects of this combination rule and range sampling, we compare three codebooks with the same angular grid, polarization structure, and power constraint. \(\mathcal C_{\mathrm{T1}}^{(\nu)}\) is the rank-\(\nu\) Type-I candidate set. \(\mathcal C_{\mathrm{PW}}^{(\nu)}\) retains the oversampled plane-wave atoms and polarization co-phases of Type-I but removes the fixed interlayer beam offsets, allowing the \(\nu\) layers to select distinct plane-wave atoms from the full angular grid. \(\mathcal C_{\mathrm{PD}}^{(\Delta,\nu)}\) retains the same free multilayer combination and adds finite-range spherical-wave atoms at every direction. They satisfy \(\mathcal C_{\mathrm{T1}}^{(\nu)}\subseteq\mathcal C_{\mathrm{PW}}^{(\nu)}\subseteq\mathcal C_{\mathrm{PD}}^{(\Delta,\nu)}\). The first expansion therefore changes only the multilayer beam combination, and the second adds only range states.

For an NF channel, define
\begin{subequations}
\begin{align}
G_{\mathrm{struct}}
&=R_{\mathrm{NF}}(\mathcal C_{\mathrm{PW}})
-R_{\mathrm{NF}}(\mathcal C_{\mathrm{T1}}),\\
G_{\mathrm{range}}^{(\Delta)}
&=R_{\mathrm{NF}}(\mathcal C_{\mathrm{PD}}^{(\Delta)})
-R_{\mathrm{NF}}(\mathcal C_{\mathrm{PW}}),\\
G_{\mathrm{total}}^{(\Delta)}
&=R_{\mathrm{NF}}(\mathcal C_{\mathrm{PD}}^{(\Delta)})
-R_{\mathrm{NF}}(\mathcal C_{\mathrm{T1}})
=G_{\mathrm{struct}}+G_{\mathrm{range}}^{(\Delta)}.
\end{align}
\label{eq:attribution}
\end{subequations}
where \(R_{\mathrm{NF}}(\mathcal C)=\max_{\mathbf W\in\mathcal C}R_\nu(\mathbf H_{\mathrm{NF}},\mathbf W)\). The three \(G_q\) terms are the absolute rate increments from the two successive changes and from their combination. Using the same SVD-EP denominator as in \eqref{eq:svd-normalized-rate}, define \(\eta_q^{(\nu)}=G_q^{(\nu)}/R_{\mathrm{SVD\text{-}EP,NF}}^{(\nu)}\), where \(q\in\{\mathrm{struct},\mathrm{range},\mathrm{total}\}\).
The three metrics for the same user and rank share the SVD-EP denominator and satisfy \(\eta_{\mathrm{total}}=\eta_{\mathrm{struct}}+\eta_{\mathrm{range}}\). Equation~\eqref{eq:svd-normalized-rate} gives \(\eta_{\mathrm{struct}}=\bar R_{\mathrm{PW,NF}}-\bar R_{\mathrm{T1,NF}}\) and \(\eta_{\mathrm{range}}=\bar R_{\mathrm{PD,NF}}-\bar R_{\mathrm{PW,NF}}\), so \(\eta_q\) directly gives the increment in SVD-EP-normalized rate.

The polar dictionary adopts nonuniform range sampling~\cite{cui2022polar}:
\begin{equation}
\rho_{\ell,n}
=\frac{Z_\Delta}{\ell}(1-\Omega_n^2),
\qquad
Z_\Delta=\frac{N^2\lambda}{8\beta_\Delta^2},
\label{eq:polar-sampling}
\end{equation}
where \(\Omega_n\) is an oversampled direction cosine, \(\beta_\Delta\) corresponds to the target adjacent-atom magnitude correlation \(\Delta\), and \(\rho=\infty\) is the plane-wave state. Exact spherical-wave inner products determine the candidate ranges retained within the InH-linear user region. For an \(N\)-element half-wavelength ULA with oversampling factor \(O\), adjacent angular codewords have correlation \(\gamma_\Omega=\sin(\pi/O)/\{N\sin[\pi/(NO)]\}\).
With \(N=80\) and \(O=4\), \(\gamma_\Omega=0.90033\); hence \(\Delta=0.90\) is the resolution-matched baseline and \(\Delta=0.95\) the denser check.

\begin{table}[!t]
\caption{InH-Linear 24-GHz Polar-Dictionary Range States}
\label{tab:polar}
\centering
\footnotesize
\setlength{\tabcolsep}{3.2pt}
\begin{tabular}{ccccc}
\toprule
\(\Delta\) & Directions & Atoms & Active & Mean/max.\\
\midrule
0.90 & 319 & 856 & 285 & 2.683/4\\
0.95 & 319 & 1292 & 297 & 4.050/6\\
\bottomrule
\end{tabular}
\end{table}

The grid contains 320 plane-wave states. Including \(\rho=\infty\), the maximum state counts per direction are five and seven for \(\Delta=0.90\) and 0.95, respectively, both fitting a 3-bit range field. With 9 angular, 3 range, and 2 polarization-co-phase bits per layer plus RI, LI, and CQI, the rank-1--4 payloads are 20, 35, 50, and 64 bits.

\subsection{Rate Gains from Multilayer Combination and Range Sampling}

Fig.~\ref{fig:attribution} shows the three rate increments over ranks~1--4 and over the rank-4 distance groups. Every box is normalized by the matched NF SVD-EP rate for the same user and rank; the black total gain is the user-wise sum of the blue multilayer-combination gain and orange range-sampling gain.

The multilayer-combination gain grows with rank, from zero at rank~1 to 20.7\% at rank~4; the rank-4 range-sampling and total gains are only 0.579\% and 21.3\%, respectively. Type-I's fixed interlayer offsets therefore account for the main recoverable high-rank SVD-EP gap.

At rank~4, the mean NF rate of Type-I is 27.04~bit/s/Hz. Relaxing the multilayer-combination constraint adds 8.28~bit/s/Hz, bringing the free-plane-wave combination to 35.32~bit/s/Hz. Finite range states then add 0.239~bit/s/Hz, bringing the polar codebook to 35.56~bit/s/Hz, close to the 36.02~bit/s/Hz of eType-II \(L=4\).

The multilayer-combination gain remains dominant in every distance group. Increasing \(\Delta\) to 0.95 adds only 0.0350~bit/s/Hz on average, showing that denser range sampling provides little further improvement.

\begin{figure*}[!t]
    \centering
    \includegraphics[width=\textwidth]{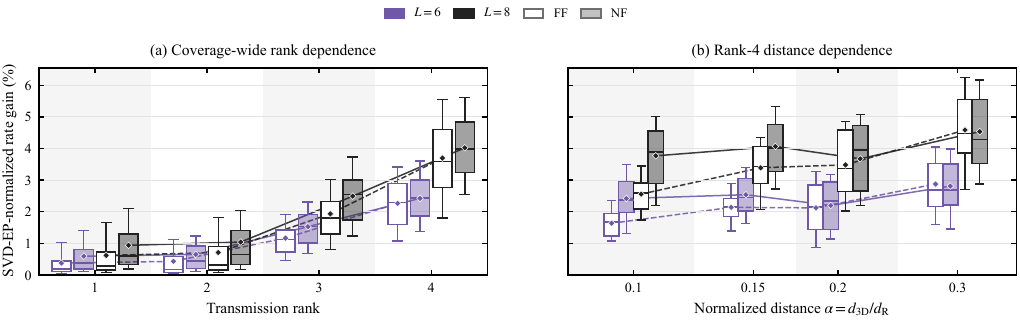}
    \caption{SVD-EP-normalized rate gains of eType-II \(L=6\) and 8 relative to \(L=4\): paired FF/NF rank dependence and paired FF/NF distance dependence at rank~4.}
    \label{fig:multibeam-extension}
\end{figure*}

\subsection{eType-II Multi-Beam Extension}

A second control keeps the eType-II plane-wave atoms, common oversampled subgrid, nonzero-coefficient selection bitmap, and amplitude/phase quantization unchanged while increasing the number of basis beams from \(L=4\) to 6 and 8. Let the absolute rate increment be \(G_X^{(L)}=R_X(\mathcal C_L)-R_X(\mathcal C_4)\). Normalizing by the SVD-EP rate for the same user, rank, and channel gives \(\eta_X^{(L)}=G_X^{(L)}/R_{\mathrm{SVD\text{-}EP},X}^{(\nu)}\) for \(X\in\{\mathrm{FF},\mathrm{NF}\}\).
Fig.~\ref{fig:multibeam-extension}(a) compares the paired FF/NF rate increments for \(L=6\) and 8. At rank~4, their mean NF normalized rate increments are 2.42\% and 4.02\%, corresponding to absolute increments of 0.964 and 1.60~bit/s/Hz; the SVD-EP gap closed by adding basis beams grows with rank.

Fig.~\ref{fig:multibeam-extension}(b) further compares the FF and NF multibeam gains within the same distance groups. At \(\alpha=0.10\), the NF normalized rate increments for \(L=6\) and 8 exceed their FF counterparts by 0.783 and 1.21 percentage points, respectively; by \(\alpha=0.30\), these FF/NF differences converge to approximately zero. Multibeam expansion is therefore effective in both FF and NF channels, while the extra gain associated with the near-field wavefront is concentrated at shorter range.

\begin{figure*}[!t]
    \centering
    \includegraphics[width=\textwidth]{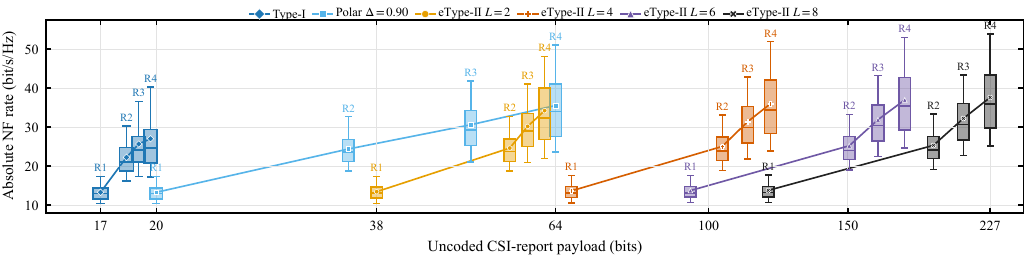}
    \caption{Rank-1--4 NF rate versus uncoded CSI-report payload for Type-I, polar \(\Delta=0.90\), and eType-II \(L=2,4,6,8\). The payload axis is logarithmic; Type-I rank-2--4 boxes are offset slightly for visibility.}
    \label{fig:feedback-tradeoff}
\end{figure*}

\subsection{Feedback--Rate Tradeoff}

The Type-I payload is accumulated as \(B_{\mathrm{CSI}}^{\mathrm{T1}}=B_{\mathrm{PMI}}^{\mathrm{T1}}+B_{\mathrm{RI}}+B_{\mathrm{LI}}+B_{\mathrm{CQI}}\). For the single-frequency evaluation, the eType-II spatial-PMI payload is
\begin{equation}
\begin{aligned}
B_{\mathrm{PMI}}^{\mathrm{eII}}(\nu,L)
={}&\left\lceil\log_2(O_1O_2)\right\rceil
+\left\lceil\log_2\binom{N_1N_2}{L}\right\rceil\\
&+B_{i_{1,8}}(\nu)+4\nu+2LM_\nu\nu+7(K_{\mathrm{NZ}}-\nu),
\end{aligned}
\label{eq:etype-feedback}
\end{equation}
The terms represent the common oversampling offset, unordered basis-beam set, strongest-coefficient indices, reference amplitudes, selection bitmap, and non-anchor coefficients; \(M_\nu=1\). RI, CQI, and the nonzero-count indicator complete the uncoded single-frequency payload; UCI coding and CRC are excluded.

Fig.~\ref{fig:feedback-tradeoff} maps the rank-1--4 NF rate distributions of Type-I, polar \(\Delta=0.90\), and eType-II \(L=2,4,6,8\) to their payloads. Its horizontal axis is logarithmic; the 19-bit Type-I rank-2--4 boxes are slightly offset for visibility.

At rank~4, Type-I and polar achieve 27.0 and 35.6~bit/s/Hz with 19 and 64 bits. eType-II \(L=2,4,6,8\) achieves 34.2, 36.0, 37.0, and 37.6~bit/s/Hz with 62, 120, 177, and 227 bits. At similar payload, polar exceeds \(L=2\) by 1.3~bit/s/Hz. The \(L=2\!\to\!4\), \(4\!\to\!6\), and \(6\!\to\!8\) steps add 58, 57, and 50 bits for 1.8, 0.97, and 0.63~bit/s/Hz, respectively, revealing diminishing returns.

Figs.~\ref{fig:attribution}--\ref{fig:feedback-tradeoff} jointly show that relaxing Type-I's fixed interlayer offsets provides the main rate increase. Once the codebook can jointly represent several strong directions, adding range states or further eType-II basis beams still improves rate, but the marginal return per feedback bit decreases.

\section{Conclusion}
\label{sec:conclusion}

This paper evaluated finite-feedback multilayer precoding using NR Type-I and eType-II CSI codebooks over TR~38.901 Release~19 near-field channels. LoS geometric analysis shows that the small UE aperture limits the formation of additional strong LoS modes in the configurations considered. Paired multipath results show no systematic FF/NF displacement in the singular-mode gains or SVD-EP rates, although individual users and modes may strengthen or weaken and spherical-wave phases reorder multipath projections onto plane-wave beams. Applying the FF-selected codeword directly to the NF channel measures the codebook's direct sensitivity to the wavefront change; after NF-based reselection, other strong multipath components absorb part of the loss. The final change is jointly determined by the multipath channel response and the codebook's adaptation capability.

The high-rank results further show that eType-II's multiple basis beams and complex weights give it a higher absolute rate than Type-I. With the angular grid, polarization structure, and power constraint fixed, relaxing Type-I's fixed interlayer beam offsets provides the main gain, whereas finite range states yield only a small additional improvement. Increasing the eType-II basis-beam count improves both FF and NF rates, but its marginal return decreases as the feedback payload grows. Large-port NR CSI codebooks should therefore prioritize multilayer multibeam combination and complex-weight capability, using range states to compensate residual wavefront-curvature mismatch.

\vspace{-1.0\baselineskip}
\renewcommand{\IEEEbibitemsep}{-0.60pt}
\bibliographystyle{IEEEtran}
\bibliography{bib}

\end{document}